\documentstyle[12pt]{article}
\newcommand{\be}{\begin{equation}}
\newcommand{\ee}{\end{equation}}
\begin{document}
\begin{titlepage}
\hoffset=-37pt
\title{Averaging, Renormalization Group and\\
       criticality in Cosmology\thanks{Talk given at the South
African Relativity Workshop, Cape Town, June 1995.}}

\author{{\sc Kamilla Piotrkowska\thanks{Present address: University of
Cape Town }}
 \\
\\
\normalsize{\it Department of Math. and Applied Math.}
\\
\normalsize{\it University of Cape Town, Rondebosch 7700,}
\\
\normalsize{\it Cape Town, South Africa.}
\\
\\
\normalsize{\it S.I.S.S.A. Strada Costiera 11}
\\
\normalsize{\it Trieste 34014, Italy.}
\\  \\
       }
\date{$\mbox{}$ \vspace*{0.3truecm} \\ \normalsize{\today}           }
\maketitle
\thispagestyle{empty}
\vspace*{0.5truecm}
\begin{abstract}
Several problems in physics, in particular the averaging problem in
gravity, can be described in a formalism derived from the real-space
Renormalization Group (RG) methods. It is shown that the RG flow is
provided by the Ricci-Hamilton equations which are thereby provided with a {\it
physical} interpretation. The connection between
a manifold deformation according to these equations and Thurston's
conjecture is exhibited. The significance of criticality which
naturally appears in this framework is briefly discussed.
This article summarizes also recent work with M.
Carfora \cite{Kami-mauro}.
Moreover, a report on some work in progress is given and some open
issues in the averaging problem pointed out.

\end{abstract}

\centerline{U.C.T. ref.}

\end{titlepage}
%
%

Any theory, a physical theory in particular, describes certain kind of
objects and their environments, e.g. matter, fields, space-times.
Among them there is a class of
objects which are easy to study, let us call them the homogeneous objects,
where what we have in mind is generally that they are highly symmetric.
The difficulties arise when we begin to examine inhomogeneous objects.

In physically realistic cases one necessarily has to deal with
inhomogeneous objects.
Homogeneous objects are easy to analyze but they are so special that it is
unlikely that a physical object of a certain kind is going to be a
homogeneous one. Thus we have to resort to studying inhomogeneous
objects despite the related mathematical difficulties arising.
Fortunately, there exist some techniques which allow us to relate various
inhomogeneous objects with homogeneous ones, in such a way that we
keep control of change in their properties and phenomena we are interested
in.

To analyze  a general case of an inhomogeneous object there are two
strategies we may adopt which we will refer to as an {\it infinitesimal
change strategy} and a {\it finite change} one.

In the first case, we have in mind a situation where, in a space of
objects, there is a special point representing a homogeneous object.
Let us perturb it slightly to an infinitesimally close position
in the tangent space of the original point.
What we get is an object which is almost homogeneous
in the sense that the inhomogeneities generated are infinitesimally small.
This kind of object is a concern of perturbation theory and this is a
powerful method in a great many cases despite its obvious limitations.
In the other possible approach, the strategy of finite change, we study
how to deform a generic object (at some finite distance from the special
one) into a special (homogeneous) one, keeping control of its properties
which change during this process.

To concretize this consideration let us concentrate on three examples.
\begin{enumerate}
\item 3-d geometry

The {\bf aim} is here to classify all 3-d manifolds. This is a difficult
and not yet achieved task. {\bf Special objects} in this context are e.g.
homogeneous manifolds. The {\bf problems} are here related
with Thurston's 3-d geometrization conjecture \cite{Thurston} which claims
that any closed three-manifold may be canonically decomposed into pieces
such that each of the pieces admits a locally homogeneous geometry. In
this case it is not clear how perturbation theory applies (strategy of
infinitesimal change) since the problems encountered are that of global
topology. A {\bf possible solution}, or rather a program (not yet fully
accomplished) for proving this conjecture was put forward by R. Hamilton.
Roughly, this program rests on the application of the Ricci-Hamilton flow,
namely, the idea is to choose an arbitrary metric on the three-manifold
and then deform it via the Ricci-Hamilton flow equation
\be
\frac{\partial g_{ab}}{\partial \eta}=\frac{2}{3}<R(\eta)>g_{ab}-
2R_{ab}(\eta),
\ee
where $<R>$ denotes the average of the scalar curvature $R$ over the
manifold and $R_{ab}$ the components of the Ricci tensor, respectively,
whereas $\eta$ is some deformation parameter.
The hope is then to relate the local singularities of the flow to the
manifold decomposition in Thurston's conjecture and to show that the
Ricci-Hamilton flow of the geometry away from each of the local
singularities approaches that of a locally homogeneous geometry in each
disconnected piece. It has been shown that for certain classes of
three-metrics the Ricci-Hamilton flow converges \cite{Hamilton, Ca-Is-Ja,
Is-Ja}.

\item Critical phenomena

Critical points (second-order phase transitions) occur in liquid-gas
transitions, ferromagnetic transitions, binary alloys, etc. There are
close analogies relating all these critical points, which is one of the
fascinations of the subject. In the case of a ferromagnet the critical
temperature $T_c$ marks the onset of spontaneous magnetization (with no
external field applied) which varies as $(T_c-
T)^{\beta}$ from below near $T_c$. The exponent $\beta$ is an example of
 a critical exponent. One
of the {\bf aims} of a theory of critical phenomena is to predict the
values of the exponents (see e.g. \cite{Binney}).
(There are other critical exponents
characterizing other power laws near the critical point). In this example
we may identify as {\bf special objects} systems far from criticality
which, for the case of a ferromagnet,  are those in low or high
temperature regimes.

Critical phenomena belong to a class of phenomena where power-law
behaviour occurs, but the exponent is not a simple fraction. This is one
of the many {\bf problems} a theory of critical phenomena must address.
It is well known that the mean-field-type-theories
(van der Waals, Weiss, Landau) lead to
exponents in form of rational fractions, in disagreement with experiment.
It happens often that apparently different physical systems exhibit
precisely the same critical exponents and this fact,
paradoxical on its face value,
 is known as universality leading to universality classes. It has been
found in scattering experiments that at the critical point large-scale
fluctuations are not exponentially rare, as they are above or below $T_c$,
and large-scale (macroscopic) dynamical structures, roughly scale-free,
are generated. This phenomenon occurs in several physical systems not
adequately understood (distribution of
earthquakes, turbulence, polymers). In each case in a wide range of
scales some phenomenon varies as a power law of the scale, presumably
because  of a gross mismatch between the largest and smallest scales
in the problem. With a {\bf solution} to these problems a new way of looking
at physics emerged, namely Renormalization Group (RG) theory of phase
transitions and critical phenomena which constitutes one of the greatest
achievements in theoretical physics of the last decade. In recognition of
this K. Wilson, who put forward the RG approach, was awarded the 1982
Nobel prize in physics.

\item Gravitation theory

The {\bf aim} of cosmology is to describe a realistic (lumpy) universe.
Unfortunately, we do not know any solutions of Einstein equations capable
of describing a lumpy universe.
For this reason, when considering the kinematics and dynamics of the
universe as a whole, one usually ignores the fine-graining due to the
local inhomogeneities and deals with the simpler structure of space-time
geometry, which is more illuminating from the point of view of cosmology.
The relation however between ``background
models" of the universe (usually taken to be homogeneous
 and isotropic) and the fine-grained
(more detailed) ones, such as to give a mathematical formulation
of the idea that the universe is on average homogeneous
is a difficult problem
(see e.g. \cite{Ellis-aver}).
This problem, known generally as the {\it averaging problem},
lead to the assumption
that, on large scales, the universe may be described by the homogeneous
and isotropic FLRW metrics, the {\bf special objects} in this context.

To investigate this issue, perturbation theory does not seem to be adequate
since the problem is generically non-perturbative in its nature.
In practical terms the heuristical justification for using FLRW models
asserts that for them to hold the matter inhomogeneities have to be
averaged (or smoothed-out) and redistributed homogeneously (e.g. in the
form of a perfect fluid). A series of related mathematical {\bf problems}
arise.
\end{enumerate}

First of all, let us notice that we are using continuous functions
in modeling the universe (matter density, pressure or kinematical scalars
of the velocity field), assuming that they represent ``volume averages"
of the corresponding fine-scale quantities. Einstein equations are solved
with a smooth (continuously distributed) stress tensor which implies
a space-time (or ensemble) averaging of a discrete matter distribution
in forms of stars, galaxies, etc, has been carried out. The results
of such averaging in an inhomogeneous medium depend on scale but this
scale was never explicitly agreed upon. Additional problem is that
a volume average for tensors is a non-covariant quantity (unlike for
scalars), so a more sophisticated definition is required. The basic
and tacit assumption which underlies this whole procedure is that a
smoothed-out universe and the actual, inhomogeneous one, behave
identically under their own gravitation. Or, more precisely, almost
identically on some scales of interest, e.g.  the ones that are much
greater than a characteristic scale of the local inhomogeneities
and much smaller than a characteristic length of the universe model
under study. This assumption is usually  taken for granted, but
does by no means need to be true. In particular, it should be stressed,
that we face here a severe problem of the non-commutativity of averaging
of the metric and calculating the Einstein tensor (non-linear in the
metric). Because of this feature averaging of the Einstein equations
is likely to be not an easy task.
Another problem of interest is here concerned with a measure of closeness
between the same physical systems  (universes) containing geometrical
information at different scales of description enabling us to resolve
less or more details \cite{Kami-spec}. Gromov's distance (see below)
might also serve as a useful concept in this context.
\bigskip

Various approaches have been attempted in tackling the averaging problem
with various competing definitions of averaging or, sometimes, with no
definition at all where the averages were introduced formally (for a
review see \cite{Kras} or \cite{Kami-phd}).
Worth mentioning is here a macroscopic gravity theory \cite{Zala}.
This is an axiomatic approach whose averaging refers to a derivation
of the so-called macroscopic field theories whereby the Einstein equations
are averaged in a covariant procedure using bilocal tensor-valued p-forms.
This approach generalizes
an averaging procedure due to Lorentz in electrodynamics in deriving
Maxwell's equations (which is a macroscopic theory of electromagnetism)
by means of a space-time averaging of the Maxwell-Lorentz equations.
For more details on the physical status of the macroscopic gravity theory
and its comparison with other approaches we refer the reader to
\cite{Kami-Zala}.
\bigskip

Now, it is the subject of this article to briefly describe another
{\bf possible approach} to tackle the averaging problem
in gravitational theory with an application to cosmology in mind
(for details we point the reader to \cite{Kami-mauro}).
This approach belongs to the finite change strategy methods remarked earlier.
Our idea was to use RG method, theory and philosophy, interpreted broadly to
include various kinds of ``multi-length-scale" and ``coarse-graining"
phenomena and arguments. This enabled us to find an interesting and deep
connection of the averaging problem with the problems of 3-d geometry, as
summarized in our first example.

The usual RG transformations invoking averaging over square blocks are
designed mostly having ferromagnetic systems in mind. However there are many
open problems suitable for the RG methods. They belong to the most difficult
problems known in physics where their difficulty can be traced to a
multiplicity of scales and where the reductionist approach fails. For each new
physical situation one has to  ``custom-make" the RG map and according to M.
Fisher: {\it A ``good" renormalization group must be  ``apt" or appropriate for
the problem at hand, and it must, in particular, ``focus" properly on the
critical phenomena of interest} \cite{Fisher}.
The RG can be used in various formulations. Apart from displaying the scale
dependence of the fundamental laws and constants, it can be useful to eliminate
unimportant variables. Usually, the mathematical expressions are similar but
the interpretations can differ.

The basic physical insight on which the technique of RG is
built, due to Kadanoff, was an assumption that near
a critical point the system ``looks the
same on all length scales" (roughly), namely, he hypothesized that
if the block lattice is considered, then at the critical point the
block-lattice Hamiltonian may be reduced to the initial lattice Hamiltonian by
scaling transformations. Then in the 70's Wilson gave a formulation which
provided a systematic way of implementing the integration over a finite
fraction of degrees of freedom in a near-critical system, and quantifying the
effect on the remaining variables, providing in this way all the mathematical
infrastructure required to explain scaling
and universality. In other words, the effect
of the long wave-length fluctuations could be calculated using self-similarity
properties of a critical system under scale transformations. When the effect
on the long wave components of the integration over the short wavelength ones
vanishes, the transformation has a fixed point which determines the universal
critical singularities. If we think of a RG transformation as a transformation
in a space of coupling constants of a theory, then the fixed point does not
flow under RG transformation. The critical behaviour of the model
can be gotten from the RG flow. The otherwise complicated flow pattern becomes
particularly simple in the vicinity of the fixed point where the linearized RG
flow and scaling hold.

It should be clear that the averaging problem in cosmology belongs to a
multi-length-scale class of problems as it is effectively a question of how a
system behaves when we make the scale successively coarser.
As such it is then  most naturally addressed using the RG approach,
thought of as a general strategy to handle problems with
multi-length-scales and enabling us to extract
the long distance behaviour of the
system. Hopefully choosing this path we will be able to say something about
scaling (possibly universality) in the universe, which RG theory neatly
explains in most cases. Indeed notice that there are
scaling laws exhibited in the
universe (e.g. the power law behaviour of the two-point correlation function
for galaxies)
which seem to be a hint that the universe might in fact be not far from
criticality. For other problems in cosmology and astrophysics where
critical phenomena of various types may play a role, see the review by
Smolin \cite{Lee}.
This being the motivation, the main question then to address is how to
implement  the RG philosophy in gravitational physics.
This is what we are going to describe now.

We can approach the problem by setting up a program for approximating
cosmological space-time solutions of Einstein's equations via the development
of a procedure for smoothing sets of initial data for such space-times. In
doing so we choose to handle the unwanted fluctuations of matter and
space-time geometry on small scales at the level of data sets, since the time
evolution of the initial data set for the Einstein equations is actually
determined by the constraints which that data set has to satisfy. Thus our
purpose is to extract {\it effective} dynamics capturing the global dynamics
of the original space-time.

The approach we adopt is that of a 3+1 formulation of GR. Let us assume that we
have a differentiable, compact riemannian three-manifold $\Sigma$ without a
boundary, to be thought of as a particular hypersurface in a 4-dim space-time.
We also assume that $\Sigma$ is a manifold of {\it bounded geometry}. Such
manifolds, or more precisely the corresponding riemannian structures, can then
be classified according to how they can be covered by small metric balls (to
be defined later). Moreover, this set of riemannian structures has some
remarkable compactness properties. This is a classical result due to M.
Gromov, related to the possibility of introducing a distance function which,
roughly speaking, enables one to say how close riemannian manifolds are to
each other. Of particular interest in this context is the fact that nearby
riemannian manifolds (in the sense of Gromov distance) can be covered with
metric balls arranged in similar packing configurations \cite{Mauro-balls,
Grove-Pet}.

To study local properties on a manifold we need to introduce coordinates. This
is an arbitrary choice but there is a preferred system of coordinates
particularly well suited to studying riemannian geometry \cite{Ellis-Matra}.
This is  the
geodesic system. This coordinate system is also a natural one from the point
of view of a human observer. The insight here is that by using, in an
appropriate way, the geodesic system of coordinates one can also give a good
description of the {\it global} properties of riemannian structures captured
by the coverings of the manifold with the metric balls.

In order to define such coverings \cite{Grove-Pet}, let us
parameterize the geodesics by arc-length,
and for any point $p\in \Sigma$, let $d_\Sigma (x,p)$ denote the
distance function of a generic point $x$ from the chosen one $p$. Then for
any given $\epsilon > 0$ it is always possible to  find an ordered set of
points $\{p_1, \cdots, p_N\}$ in $\Sigma$, so that

\begin{description}
\item the open metric balls (geodesic balls)
$B_\Sigma(p_i,\epsilon)=\{x\in\Sigma|\,d_\Sigma(x,p_i)<\epsilon\},\,i=1,\cdots,
N$, cover $\Sigma$; in other words the collection $\{p_1,\cdots,p_N\}$ is an
$\epsilon$-net in $\Sigma$.
\item  the open balls $B_\Sigma(p_i, \epsilon /2), i=1,\cdots, N$, are
disjoint,   i.e. $\{p_1,\cdots,p_N\}$ is a {\it minimal} $\epsilon$-net in
$\Sigma$.
\end{description}

This way we obtain a discretized manifold model which  is obviously simpler
to  deal with than the original manifold,
in the sense that it is a union of geodesic balls. In
particular we can carry out in this model averaging of tensors, which we cannot
in curved space. Moreover, this discrete object describes the manifold well,
the underlying construction being somewhat similar in spirit to
that of Regge Calculus
or dynamical triangulations used in Monte Carlo approaches to  4-dim Quantum
Gravity. It luckily happens also that the $\epsilon$-nets just
introduced are useful from a physical point of view.
To see this let us consider the average of
a scalar function $f$ on $\Sigma$
\be
\label{eq1}
<f>_{\Sigma(g)}=\frac{\int_{\Sigma}f d\mu_g}{vol(\Sigma,g)},
\ee
where $vol(\Sigma,g)=\int_{\Sigma} d\mu_g$ and $\mu_g$ is the riemannian
measure associated with the three-metric $g$ of $\Sigma$.
If the geometry of $\Sigma$ is not known on large scale we cannot take
(\ref{eq1}) as an operational way of defining the average of $f$. Taking a
pragmatic viewpoint and assuming that we can only experience geometry in
sufficiently small neighbourhoods of a finite set of instantaneous
observers, it makes much more sense to replace (\ref{eq1}) with a suitable
average based on the geometrical information available on the length scale
of such observers.
For simplicity, given a finite set of instantaneous observers, located at the
points $x_1,\cdots,x_N \in \Sigma$, we may assume that these regions
susceptible to
observation are suitably small geodesic balls of radius $\epsilon$,
scattered over the hypersurface $\Sigma$ so as to cover it. In other words,
we assume that $\{x_1,\cdots, x_N\}$ is a minimal $\epsilon$-net in $\Sigma$.
With the above in mind, we can approximate the average of $f$ over $\Sigma$ by
averaging over euclidean balls, whose Lebesgue measure is locally weighted  by
Puiseux' formula, rewriting formula (\ref{eq1}) to leading order as
\be
\label{eq2}
<f>_{\Sigma(g)}\simeq <f>_\epsilon \equiv \frac{\Sigma_i [f_i+(\frac{\Delta
f_i - R_i f_i/3}{2(n+2)}) \epsilon^2]}{\Sigma_i[1-\frac{R_i}{6(n+2)}
\epsilon^2]},
\ee
where $f \equiv f(x_i),\,\Delta$ is the Laplacian operator relative to the
manifold, $R_i\equiv R(x_i)$ is the scalar curvature at the center of the
ball and $n={\rm dim}\,\Sigma$. This formula (\ref{eq2}) can then be taken
as a suitable scale dependent
approximation to $<f>_{\Sigma(g)}$.

There are certain problems here which need to be clarified. Obviously, there
are ``unwanted" details affecting this averaging, the immediate one being that
associated with use of a collection of geodesic balls.
The important question to ask
is what happens to the average $<f>_\epsilon$ when we change the length scale
represented by the radius of the balls. Clearly, on scales big enough the
average should not be sensitive to the details of the underlying   geometry
since homogeneity and isotropy prevail. This is the reason why averaging
in constant
curvature spaces is well defined since there one can move the balls
freely and deform them but by so doing no new geometric details, which measure
the inhomogeneities, will be felt in the average values. A natural question to
ask is then how the geometry, i.e. curvature inhomogeneities, should depend on
scale so that the average $<f>_\epsilon$ over the balls is scale independent,
or equivalently, how do we have to deform the geometry in order to achieve the
scaling limit when the size of the balls no longer matters?

To this end, let us consider the average $<f>_{\epsilon_o +\eta}$, with $\eta$
a positive number, $\eta/\epsilon_o \ll 1$. Upon expanding
$<f>_{\epsilon_o+\eta}$ in $\eta$ we get to leading order
\be
\label{eq3}
\epsilon_o\frac{d}{d\eta } <f>_{\epsilon_o + \eta}|_{\eta/\epsilon_o=0}=
\frac{1}{n+2}<\Delta f>_{\epsilon_o}+\frac{1}{3(n+2)}[<R>_{\epsilon_o}
<f>_{\epsilon_o}-<Rf>_{\epsilon_o}],
\ee
where $<f>_{\epsilon_o}$  is the average of $f$ over the set of $N$
instantaneous observers (with similar expressions for $<R>_{\epsilon_o}$,
$<Rf>_{\epsilon_o}$ and $<\Delta f>_{\epsilon_o}$).
It is evident from the above formula that on a curved,
inhomogeneous, manifold the curvature
necessarily enters in the process of averaging. Namely, note the following
feature of (\ref{eq3}): if $R={\rm const}$
the second term on the r.h.s. of (\ref{eq3}) disappears. In these
circumstances, if $f$ is related to geometry, one cannot speak of averaging
without taking into account the backreaction of the curvature.

Since, in general, sources are coupled to the gravitational field, we ought to
treat the full dynamical system, i.e. the cosmological fluid and the geometry
of $\Sigma$, in the RG procedure of blocking. In this setting the blocking is
to
be understood as passing from an $\epsilon_m$-net to an
$\epsilon_{m+1}$-net, given
by a collection of geodesic balls with radius $\epsilon_m \equiv
(m+1)\epsilon_o$ for $m=0,1,2,\cdots$, thought of as  \`{a} la
Kadanoff blocking, applied here to the geometry itself.
In the RG treatment of the
lattice models in statistical mechanics, blocking
engenders a change (flow) of the coupling constants in
the corresponding Hamiltonian, according to the RG equations. We wish to
imitate the above picture in gravity and thus blocking of the manifold,
in our setting, should give rise to a deformation of
geometry. Consequently one can see here that it is not possible to provide a
consistent averaging procedure of the sources without taking into account the
backreaction of the geometry.

The next step to consider is then a deformation of geometry.
How do we achieve this? Let us formally write that the effect
of the renormalization induced by the
blocking (and appropriate rescaling, see \cite{Kami-mauro} for details)
can be symbolized as a non-linear
operation acting on a metric $g^{(m)}$  so as to produce $g^{(m+1)}$, i.e.
\be
\label{formal}
g^{(m+1)} ={\cal R} (g^{(m)}).
\ee
Since it is natural in this setting to consider the metric as a running
coupling constant, depending on the cut-off, while thinking of the average
$<f>_\epsilon$ as a functional of the metric, we can equivalently interpret
(\ref{eq3}) as obtained by considering the variation of $<f>_\epsilon$ under a
suitable smooth deformation of the metric, rather than deforming the
(euclidean) radius of the balls $\{B(x_i,\epsilon)\}$. We can equivalently
rewrite the second term on the r.h.s. of (\ref{eq3}) as
\be
<R>_\epsilon <f>_\epsilon -<Rf>_\epsilon = -D<f>_\epsilon \cdot
\frac{\partial g_{ab}}{\partial \eta},
\ee
where $D<f>_\epsilon \cdot \partial g_{ab}/\partial \eta$ denotes a formal
linearization of the functional $<f>_\epsilon$ in the direction of the
symmetric 2-tensor $\partial g_{ab}/\partial \eta$, and where
\be
\label{eq4}
\frac{\partial g_{ab}(\eta)}{\partial \eta}=\frac{2}{3}<R(\eta)>g_{ab}(\eta) -
2 R_{ab}(\eta),
\ee
$R_{ab}(\eta)$ being the components  of the Ricci tensor and $<R(\eta)>$ is
the average scalar curvature given by
\be
<R(\eta)>=\frac{1}{vol\,\Sigma}\int_\Sigma R(\eta) d\mu_\eta.
\ee
The metric flow (\ref{eq4}) is known as the Ricci-Hamilton flow
studied in connection with the quasi-parabolic flows on manifolds.
Its use in cosmology advocated earlier in  \cite{Mauro-Hflow} was
quite {\it ad hoc}. The Ricci-Hamilton flow is always solvable for sufficiently
small $\eta$ \cite{Hamilton} and
has a number of useful properties, namely, it preserves the volume (due to
the normalization chosen), and any symmetries of the original metric are
preserved along the flow, and the limiting metric $\bar{g}_{ab} = \lim_{\eta
\rightarrow \infty} g_{ab}(\eta)$ (if attained) has constant sectional
curvature.

Smoothing-out matter sources, as described by a set of instantaneous observers,
means thus eliminating from the distribution of such sources on $\Sigma$ all
fluctuations on scales smaller than the cut-off distance $\epsilon$, leaving
an effective probability distribution of fluctuations for the remaining
degrees of freedom. The underlying RG philosophy tells us that this effective
distribution has the same properties as the original one at distances much
larger than $\epsilon$ (i.e. for fluctuations  with wavelengths much larger
than $\epsilon$). Since we are adopting the Hamiltonian point of view, the
field $f$ characterizing matter sources, as described by the instantaneous
observers on $\Sigma$ is given by
\be
f\equiv \alpha \rho + \alpha^i J_i,
\ee
where $\rho$ is the matter density, ${\bf J}$ momentum density and $\alpha,\,
\alpha^i$ stand for the lapse function and shift vector, respectively, whose
assignment on $\Sigma$ specifies the observers in question. The independent
parameters in the effective Hamiltonian $H^{(m)}$ are then the metric
and the second fundamental form $K$ (whose renormalization is generated
by the linearization of (\ref{formal})), whereas
$\rho^{(m)}$ and ${\bf J}^{(m)}$ are at each stage of the renormalization
connected to $g^{(m)}$ and $K^{(m)}$ by the constraints that hold at each
stage (they fix the effective Hamiltonians which otherwise are undetermined up
to a constant factor \cite{Kami-mauro}).
These requirements imply that the full effective
Hamiltonian (matter and geometry) takes on the standard ADM form pertaining
to gravity interacting with a barotropic  fluid at each stage of
renormalization. This way the renormalization of matter fields is
intrinsically tied  with the renormalization (deformation) of the metric.

The invariance of long distance properties of the matter distribution, under
simultaneous change of the cut-off (blocking) and deformation of $g$, can be
expressed  as a differential equation for the effective Hamiltonian ${\cal H}
(\rho, {\bf J})$ (actually for its partition function), namely,
\be
\label{eq5}
[-\epsilon \frac{\partial}{\partial \epsilon}+\beta_{ab} (g)
\frac{\partial}{\partial g_{ab}}] \sum_{\rho, {\bf J}} \exp[-{\cal H} (\rho,
{\bf J})]=0,
\ee
where $\beta_{ab}(g) \equiv \epsilon \frac{\partial}{\partial \epsilon}
 g_{ab}$ is the $\beta$-function associated with (\ref{eq5}).
It can be shown that in order that (\ref{eq5}) is satisfied the
$\beta$-function has to be
given  by (\ref{eq4}), where the parameter $\eta$ is the
logarithmic change of the cut-off length $\epsilon$.
We have thus arrived at the conclusion that in order to reach
 a fixed point of (\ref{eq3}) the
geometry has to be deformed according to the Ricci-Hamilton flow (\ref{eq4}).

There are sets of metrics flowing under (\ref{eq4}) to respective limit
points. All those limit points, either fixed or not, have their own basins of
attractions. There are 9 such basins, corresponding to the classes of
homogeneous geometries that can be used to model the locally inhomogeneous
geometries on closed three-manifolds (cf. Thurston's conjecture).
It is here where Hamilton's program, which is an analytical approach to
Thurston's geometrization conjecture, enters the stage. This program,
roughly speaking,  amounts to saying
that any three-manifold can be decomposed into pieces on which the
Ricci-Hamilton flow is global and thereby  each
of those  pieces can be smoothly
deformed into a  locally homogeneous three-manifold.
Hamilton noticed that there are patterns in the kind of singularities that may
develop in the regions connecting the smooth-able pieces \cite{Ca-Is-Ja,
Is-Ja}.

The above analysis of the averaging problem in
cosmology can be seen as
a physically non-trivial application of the Hamilton-Thurston
geometrization programme. It is striking that motivations coming from geometry
and physics, namely, the construction of cosmological models out of a local
gravitational theory, go hand in hand in such a way.

Considering further the Ricci-Hamilton evolution of some model geometries,
 one can give characterization of the  critical fixed points associated
with the Ricci-Hamilton flow (\ref{eq4}) \cite{Kami-mauro}.
Since in this cases we are in the presence of significant finite-size effects,
the existence of the critical points is accompanied by the effective
reduction of dimensionality (crossover), and the stable phases,
under the renormalization
generated by (\ref{eq4}), are separated by a critical surface.
It is tempting to speculate that,
a striking feature of these topological crossover phenomena, associated with
the renormalization of the cosmological matter distribution, is that their
pattern resembles the linear sheet-like (sponge-like) structure in the
distribution of galaxies on large scales.
It happens that if the initial data set $(\Sigma, g, K, \rho, {\bf J})$ for
the real universe is close to criticality then the   corresponding averaged
model exhibits a tendency to such topological crossovers in various regions.
Filament-like and sheet-like structures would emerge, and the overall
situation would be the one where such structures appear altogether with
regions of high homogeneity and isotropy, in some sort of hierarchy. This
situation is akin to that of a ferromagnet near its critical temperature,
whereby we have islands of spins up and down in some sort of nested pattern.
It is fair to say that an ultimate solution to the averaging problem is
in this picture
connected to the understanding of the critical surface in the space of
riemannian metrics. Understanding this surface is however equivalent, roughly
speaking, to proving Thurston's conjecture which is still one of
the grandest topics of research in mathematics.
\bigskip

In this way the above analysis makes accessible in cosmology too the whole
subject of critical phenomena. They are manifested macroscopically since
phase transitions are collective phenomena. To gain a thorough
understanding of their relevance in structure formation (possibly pattern
formation as well) and clustering in the universe will however require
further research to be done.
It is worthwhile to notice that the notion of criticality is of particular
importance in non-perturbative formulations of Quantum Gravity where the
problem of the classical limit seems to be exactly that of critical phenomena,
namely, the theory has to be tuned to a critical point to achieve this
limit (see e.g. \cite{LeeQG} for a review).
In this sense a classical space-time emerges in a limit somewhat similar
to a thermodynamical limit. Perhaps this suggests a deep connection between
space-time, thermodynamics and criticality. The numerical results on black
hole gravitational collapse (e.g. \cite{Choptuik}) seem to point in this
direction. We venture to suggest that the critical exponent found in these
simulations, close to 0.37, for the black hole mass is a deep property of the
gravitational
field equations.
It thus seems likely that critical phenomena found in gravitational collapse
could
be described using RG approach, treating Einstein's equations as generating
a RG flow on the space of initial data (for an attempt in this direction
see \cite{Koike}).
Independently, we notice that the above  connection is already
at work  in the evaporation of black holes.

\bigskip
{\bf Acknowledgements}

\noindent
I benefited from discussions with J. P\'{e}rez-Mercader, R. Zalaletdinov,
J. Devoto, P. Haines, M. Carfora and D. Coule. I thank G. Ellis for helpful
comments and encouragement. I also thank D. Sciama for hospitality at SISSA.

\noindent
F.R.D. (South Africa) is acknowledged for financial support.


\newpage
\begin{thebibliography}{99}
\bibitem{Kami-mauro} M. Carfora, K. Piotrkowska, {\it A Renormalization Group
approach to relativistic Cosmology}, to appear in Phys. Rev. D.

\bibitem{Thurston} W. Thurston, Bull. Amer. Math. Soc. (N.S.) {\bf 6} (1982)
357.

\bibitem{Hamilton} R. Hamilton, J. Diff. Geom. {\bf 17} (1982) 255.

\bibitem{Ca-Is-Ja} M. Carfora, J. Isenberg, M. Jackson,  J. Diff. Geom.
{\bf 31} (1990) 249.

\bibitem{Is-Ja} J. Isenberg, M. Jackson, J. Diff. Geom. {\bf 35} (1992) 723.

\bibitem{Binney} J. J. Binney, N. J. Dowrick, A. J. Fisher, M. E. J. Newman,
 {\it The theory of critical phenomena. An introduction to the
Renormalisation Group}, Clarendon Press Oxford (1992).

\bibitem{Ellis-aver} G. F. R. Ellis, in Proceed. of the 10-th International
conference on GR and Gravitation, ed B. Bertotti et al, Reidel Dordrecht
(1984).

\bibitem{Kami-spec} K. Piotrkowska, {\it Spectral representation
of the space-time geometry in averaging}, work in progress.

\bibitem{Kras} A. Krasi\'{n}ski, {\it Physics in an Inhomogeneous Universe},
to be published by C.U.P.

\bibitem{Kami-phd} K. Piotrkowska, Ph.D. thesis, SISSA (1994).

\bibitem{Zala} R. M. Zalaletdinov, Gen. Rel. Grav. {\bf 24} (1992) 1015;
{\it ibid.} {\bf 25} (1993) 673.

\bibitem{Kami-Zala} K.Piotrkowska, R. M. Zalaletdinov, {\it Some issues
on averaging problem in General Relativity}, work in progress.

\bibitem{Fisher} M. E. Fisher, in {\it Critical Phenomena} Lecture Notes in
in Physics {\bf 186} ed. F. J. W. Hahne, Springer Verlag Berlin (1983).

\bibitem{Lee} L. Smolin, {\it Cosmology as a problem in critical phenomena},
gr-qc/9505022.

\bibitem{Grove-Pet} K. Grove, P. V. Petersen, Annals of Math.  {\bf 128}
(1988) 195.

\bibitem{Mauro-balls} M. Carfora, A. Marzuoli, Class. Qu. Grav. {\bf 5} (1988)
659.

\bibitem{Ellis-Matra} G. F. R. Ellis, D. Matravers, {\it General covariance
in General Relativity?}, to appear in Gen. Rel. Grav.

\bibitem{Mauro-Hflow} M. Carfora, A. Marzuoli, Phys. Rev. Lett.  {\bf 53}
(1984) 2445.

\bibitem{LeeQG} L. Smolin, {\it Experimental signatures of Quantum Gravity},
gr-qc/9503027.

\bibitem{Choptuik} M. Choptuik, in {\it Approaches to Numerical
Relativity}, ed R. d'Inverno, C. U. P. (1992).

\bibitem{Koike} T. Koike, T.Hara, S.Adachi, {\it Critical behaviour in
gravitational collapse of radiation fluid - a renormalization group
(linear perturbation) analysis}, to appear in Phys. Rev. Lett.

\end{thebibliography}
\end{document}